\newif\ifpdf
\ifx\pdfoutput\undefined
  \pdffalse              
\else
  \pdfoutput=1           
  \pdftrue
\fi
\documentclass[aps,prb,twocolumn,superscriptaddress]{revtex4}
\ifpdf
  \usepackage[pdftex,
        colorlinks=true,
        pdftitle={A New face on old code},
        pdfauthor={P.F. Peterson and Th. Proffen},
        pdfsubject={},
        pdfkeywords={NOBUGS2002/024},
        pdfpagemode=None,
        bookmarksopen=true]{hyperref}
  \usepackage{thumbpdf}
\fi
\usepackage{graphicx}
\usepackage{amssymb}

\begin{document}


\title{A New face on old code - NOBUGS2002/024}


\author{P.F. Peterson}
\affiliation{Intense Pulsed Neutron Source, Building 360, 9700 South
Cass Ave, Argonne, IL 60439-4814, USA}
\author{Th. Proffen}
\affiliation{Los Alamos National Laboratory, LANSCE-12, Mailstop H805,
Los Alamos, NM 87544, USA}


\date{October 16, 2002}

\begin{abstract}
In science we often use established code that was developed many years
ago. Frequently the documentation is lost and many of us know the
feeling of putting a number in the wrong column and trying to
understand why the program is not working. However, these old codes
are usually very useful, well tested and established. Rather than
writing new code from scratch, which must first be tested against the
established standard, it is desirable to put a new front end on the
standard, commonly referred to as 'wrapping'. This paper will discuss
wrapping of FORTRAN code using a selection of other languages to
provide simpler user interface.
\end{abstract}


\maketitle



\section{Introduction}

In the ideal world data processing and modeling techniques are
thoroughly studied and understood at the beginning of a research
program. Then someone sits down and writes easily maintainable and
extendible code with lots of comments throughout. Finally someone
writes a detailed user manual including examples and what methods were
used, then confirms that everything does indeed work on all of the
supported platforms. However, this is almost never the case.

Normally, one is trying to understand some science and has little time
to do more than write {\it ad hoc} code that frequently has a
confusing interface and only is known to work for the parameters of
the original problem. After time this type of software is grouped
together in an analysis or modeling package that is fully functional,
but lacks a real interface. Because of a steep learning curve to use
the software a new user must have lots of interaction with the author
to start to apply it to the current situation. This is a waste of
everyone's time. The solutions to this problem are to undergo
extensive rewrites of the code to simplify its user interface, or to
wrap the code with new software which deals with all of the
idiosyncrasies of the various input files. This paper will discuss the
latter option.

\section{Simple Wrapping}

Several cases of wrapping fall into a simple class where the
underlying FORTRAN runs either off of a input file or a few command
line parameters. This section will demonstrate an easy way to call
such code.

We will start with the absolute simplest program to call: {\tt hello
world}. This will be done with a {\it very} simple FORTRAN program
that requires no input or user interaction.
{\small
\begin{verbatim}
        program hello
c
        write(*,*) 'Hello world'
        end
\end{verbatim}
}
\noindent Once compiled this will have either the name {\tt hello} in
Linux or {\tt hello.exe} in DOS. For the rest of the paper we will
assume that Linux is being used. To confirm that the program works we
simply try (using a GNU compiler):
\begin{verbatim}
% g77 -o hello hello.f
% hello
Hello world
%
\end{verbatim}
\noindent Even though this example is very simple it represents the
vast majority of analysis and modeling software. The wrapping program
in these cases can write the input file and call the FORTRAN using
parameters that have already checked to avoid errors in execution.

Most require input either from a simple ASCII file (which can be set
up by the wrapping software also) or a few command line parameters and
be called in exactly the same way as our simple FORTRAN program. The
examples in this section do not go over writing input files or parsing
output files because the methods for calling the FORTRAN program is
unchanged.

\subsection{Hello Perl}

In Perl~\cite{perl;web,perlnut;bk99}, and many other languages, a
simple {\tt hello world} type of program can be executed in a few
lines. In this case the preferred function is {\tt system} and the
program is as follows (assuming that Perl is in {\tt /usr/bin})
{\small
\begin{verbatim}
#!/usr/bin/perl
$command="hello";
system($command);
\end{verbatim}
}
\noindent This can be tested on the command line
\begin{verbatim}
% hello.pl
Hello perl
%
\end{verbatim}

The {\tt system} function gives control of the current thread to the
function call and takes it back once the program executed is done. The
exit status of the program executed (multiplied by 256) is the return
value of {\tt system}. There is also a similar function, {\tt exec},
which starts the system call but never regains control. Because of the
nature of the two commands, {\tt system} is the generally preferred
method since it provides a means to step through a series of programs
in an orderly fashion.

\subsection{Hello Java}
Java~\cite{java;web,java;bk00} is a significantly more structured
language than Perl. Because of this, a simple 'hello world' program takes
several more lines, while the basics are the same.
{\small
\begin{verbatim}
import java.io.*;
class hello{
  public static void main(String args[]){
    String command="hello";
    try{
      Runtime.getRuntime().exec(command);
    }catch(java.io.IOException e){
      // let it drop on the floor
    }
  }
}
\end{verbatim}
}
\noindent Compiling and running this program results in 
\begin{verbatim}
% javac hello.java
% java hello
%
\end{verbatim}
\noindent There is no output from the executed code.

The lack of output comes from the {\tt exec} command which grabs the
input, output, and error streams from the program executed. To have
the result of our program shown on the screen we need to ask {\tt
exec} for the process' output stream so we can redirect it to the
console. This can be done by
{\small
\begin{verbatim}
...
Process process=Runtime.getRuntime().exec(command);
InputStream in_stream=process.getInputStream();
InputStreamReader in_reader=
                  new InputStreamReader(in_stream);
BufferedReader in=new BufferedReader(in_reader);
while(true){
    output=in.readLine();
    if( output==null ){
        break;
    }else{
        System.out.println(output);
    }
}
...
\end{verbatim}
}

After the redirection lines are added execution results in
\begin{verbatim}
% javac hello.java
% java hello
Hello java
%
\end{verbatim}
Much like the Perl example, this can be repeated with several different
programs to allow for stepping through an analysis process.

\section{Interactive Wrapping}

While the examples above represent the majority of wrapping needed,
there are other cases which are important to discuss. Some other tasks
are to deal with an interactive program or call FORTRAN subroutines
directly. Here we will discuss the former since the latter frequently
requires an interface layer, such as C/C++, which calls the FORTRAN
subroutines. Wrapping an interactive program happens when the original
author spent the time to make a front-end to a program but new users
want to use it through a GUI or with other programs in a seamless
manner.

\subsection{KUPLOT and Perl}

The example here is taken from PDFgetN~\cite{peter;jac00} which places
a simple graphical user interface on analysis code written in
FORTRAN. While several programs are called similar to the example
above, part of the interface is plotting the data in various stages of
the analysis. KUPLOT~\cite{kuplot;web99} is a plotting package written
using the PGPLOT library~\cite{pgplot;web94}. The three important
subroutines are to open a connection to the outside program, send
commands to it, and close the connection. Of course, if interactivity
is not needed then the methods in the previous section can be used for
any software. First we show how to open a connection to KUPLOT
{\small
\begin{verbatim}
sub kuplot_open {
  my $cmd=File::Spec->canonpath("$exepath/kuplot");
  (return)unless(-x $cmd || -x "$cmd.exe");
  my $pid = open2(\*kin, \*kout, "$cmd")
            || die ('KUPLOT connection failed ..');
  kout->autoflush();
  kin ->autoflush();
  ...
\end{verbatim}
}
\noindent The important part of this subroutine is the {\tt open2}
method. This opens reading and writing connections to the same
external code. The notation {\tt $\backslash\ast$kin} denotes that a
reference is being used to pass the handle {\tt kin}. Here {\tt kin}
is the output stream KUPLOT and {\tt kout} is input stream of
KUPLOT. In other terms, our program writes to {\tt kout} and reads
from {\tt kin} in order to hold a conversation with KUPLOT.

Next we send information to KUPLOT
{\small
\begin{verbatim}
sub kuplot_send {
  my ($cmd)=@_;
  print kout "$cmd\n";
  print kout "echo done\n";
  while(<kin>){
    (last) if(/echo done/);
  }
}
\end{verbatim}
}
\noindent A series of commands is given to KUPLOT followed by {\tt
echo done}. This is done so the output of the commands are parsed
until the {\tt echo done} is returned, which lets the calling program
know that all of the other commands were completed.

Finally we close the connection to KUPLOT
{\small
\begin{verbatim}
sub kuplot_close {
  print kout "exit\n\n";
  close(kin);
  close(kout);
}
\end{verbatim}
}
\noindent This is by far the simplest of the three subroutines. It
tells KUPLOT to execute the {\tt exit} command, then closes the input and
output streams.

\subsection{BLIND and Java}

The example in this section is a program called BLIND developed for
the Single Crystal Diffractometer (SCD) at the Intense Pulsed Neutron
Source (IPNS) at Argonne National Laboratory. Blind determines the
orientation matrix, $UB$,~\cite{busin;ac67} and lattice parameters
consistent with a given set of Bragg peaks. Here the purpose of
wrapping was to dramatically shorten the development time of
visualization and analysis software by simply using existing code and
interacting with its results. Similar to the KUPLOT example there are
three basic steps: starting a process, interacting with it, and ending
the process. The first step is fairly simple given the earlier Java
example
{\small
\begin{verbatim}
public Process startProcess(String command,
                    String cwd) throws IOException{
  // must specify a command
  if(command==null) return null;
  // get the working directory
  if(cwd!=null && cwd.length()<=0) cwd=null;
  if(cwd!=null){
    File dir=null;
    if(isDirectory(cwd)){
      dir=new File(cwd);
      return Runtime.getRuntime()
                       .exec(command,null,dir);
    }
  }else{
    // call the shorter version of the command
    return Runtime.getRuntime().exec(command);
  }
}
\end{verbatim}
}
\noindent This method allows for starting a process and specifying its
current working directory, {\tt cwd} . Without the extra parameter Java
assumes that all processes are executed from the directory where the
Java session was started.

The process of interaction is similar to the previous example where
things are read from and written to the process. Again, this method of
interaction assumes a fair amount of predictability in how the program
works. Since BLIND does not return a generic {\it done} message for
each interaction, the possible results for each step and what to do
next need to be coded into the wrapper.
{\small
\begin{verbatim}
...
InputStream  in_stream=process.getInputStream();
OutputStream out_stream=process.getOutputStream();
InputStreamReader in_reader=
                  new InputStreamReader(in_stream);
OutputStreamWriter out_write=
                new OutputStreamWriter(out_stream);
BufferedReader in=new BufferedReader(in_reader);
BufferedWriter out=new BufferedWriter(out_write);

output=this.readline(in);
while( output==null || 
       output.indexOf("Input reflection from")<0 ){
  if( output!=null && output.length()>0){
    System.out.println(output);
  }
  output=this.readline(in);
}
this.writeline(out,"y");
System.out.println(output+"y");
...    
\end{verbatim}
}
\noindent As in the simple Java example, the process input and output
must be redirected by the caller to make the program work as
expected. Lines of information are read from the process until the
proper string is seen which needs actions taken. Then the appropriate
answer is given to the process (and the console) to make it look like
the program is being automatically run by Java. The reason for writing
out the communication to the command line is to aid in debugging and
provide a means to assure the user that the program is being run as
expected.

The {\tt readline} and {\tt writeline} methods are wrapping common
functions in a convenience method. Since {\tt writeline} is
straightforward it is omitted.
{\small
\begin{verbatim}
public String readline(BufferedReader in)
                            throws IOException{
  char buff[]=new char[BUFF_SIZE];
  String result=null;
  int go_back=0;
  int size=0;

  while( ! in.ready() ){
    // wait until it is ready
  }
  for( size=0 ; size<BUFF_SIZE ; size++ ){
    if(! in.ready() ){
      if(size==0) return null;
    }else{
      buff[size]=(char)in.read();
      String charac=
        (new Character(buff[size])).toString();
      if(charac.equals("\n")){ 
        go_back++;
        break;
      }else if(buff[size]==-1){
        go_back++;
        break;
      }
    }
  }

  if(size<go_back) return "";
  result=new String(buff,0,size);
  return result;
}
\end{verbatim}
}
\noindent The reason for not using Java's default {\tt readLine}
method is that it stops reading only at the end of line
character. Since BLIND has all of its input at the end of a line
rather than on a separate line, some output is never fully written
which causes the default {\tt readLine} method to wait forever.

Finally the process is ended by a simple {\tt process.waitFor()}. The
connections to the input and output streams of the process are closed
and memory freed automatically by Java.

\section{Summary}

This paper demonstrated methods for wrapping FORTRAN code. The
techniques shown did not use any special interaction between the
wrapping code and FORTRAN program so the same methodology can be
applied to wrapping software written in any language. Most modern
languages have a method similar to the {\tt system} command of Perl
and {\tt exec} command of Java. In cases where this, in combination
with input and output files, is enough interaction between programs
wrapping is very straightforward. Also discussed was the case of
dealing with interactive sessions. While this is, in theory,
straightforward some of the idiosyncrasies were demonstrated.


\begin{acknowledgments}
Argonne National Laboratory is funded by the U.S. Department of
Energy, BES-Materials Science, under Contract W-31-109-ENG-38. Los
Alamos National Laboratory is funded by the US Department of Energy
under contract W-7405-ENG-36.
\end{acknowledgments}


\bibliographystyle{aip}

\end{document}